\documentclass{article}

\usepackage{PRIMEarxiv}

\usepackage[utf8]{inputenc} 
\usepackage[T1]{fontenc}    
\usepackage{hyperref}       
\usepackage{url}            
\usepackage{booktabs}       
\usepackage{amsfonts}       
\usepackage{nicefrac}       
\usepackage{microtype}      
\usepackage{lipsum}
\usepackage{fancyhdr}       
\usepackage{graphicx}       
\graphicspath{{media/}}     

\usepackage{array}
\usepackage{multirow}
\usepackage{csquotes}

\usepackage{booktabs} 
\usepackage{array}
\usepackage{multirow}
\usepackage[export]{adjustbox}
\usepackage{pifont}
\usepackage{subfigure}
\usepackage{enumitem}

\usepackage[normalem]{ulem}
\usepackage{setspace}

\title{What Do You Get from Turning on Your Video? Effects of Videoconferencing Affordances on Remote Class Experience During COVID-19
}

\author{
  Yanting Wu\textsuperscript{1}, 
  Yuan Sun\textsuperscript{2}, 
  S. Shyam Sundar\textsuperscript{2}
  \\
  {1} College of Information Sciences and Technology, Pennsylvania State University
  \\
  {2} The Donald P. Bellisario College of Communications, Pennsylvania State University
}

\begin{document}
\maketitle

\begin{abstract}
The outbreak of COVID-19 forced schools to swiftly transition from in-person classes to online or remote offerings, making educators and learners alike rely on online videoconferencing platforms. Platforms like \textit{Zoom} offer audio-visual channels of communication and include features that are designed to approximate the classroom experience.  However, it is not clear how students’ learning experiences are affected by affordances of the videoconferencing platforms or what underlying factors explain the differential effects of these affordances on class experiences of engagement, interaction, and satisfaction. In order to find out, we conducted two online survey studies: Study 1 (N = 176) investigated the effects of three types of videoconferencing affordances (i.e., modality, interactivity, and agency affordances) on class experience during the first two months after the transition to online learning. Results showed that usage of the three kinds of affordances was positively correlated with students' class engagement, interaction, and satisfaction. Perceived anonymity, nonverbal cues, and comfort level were found to be the key mediators. In addition, students’ usage of video cameras in class was influenced by their classmates. Study 2 (N = 256) tested the proposed relationships at a later stage of the pandemic and found similar results, thus serving as a constructive replication. This paper focuses on reporting the results of Study 1 since it captures the timely reactions from students when they first went online, and the second study plays a supplementary role in verifying Study 1 and thereby extending its external validity. Together, the two studies provide insights for instructors on how to leverage different videoconferencing affordances to enhance the virtual learning experience. Design implications for digital tools in online education are also discussed.
\end{abstract}


\section{Introduction}

In early 2020, the coronavirus pandemic started sweeping the world. Aside from being a severe public health emergency, it also disrupted the education sector by necessitating that in-person classes be transformed into technology-mediated virtual learning. Videoconferencing (VC) technology quickly became widely adopted. With teachers and students scrambling to adjust to online learning via VC platforms such as Zoom, there was little opportunity to assess how the learning experience would be affected. Two years later, it is still not quite clear how the virtual class experience is affected by the confluence of new technologies. Clarity on this topic, therefore, is important for improving pedagogical skills and strategies for online education. In the long term, this may become more pressing with the hybrid mode of education that seeks to integrate traditional in-person education with online education. Therefore, a theoretical understanding of how technologies affect learner engagement is critical before we can leverage the technology for improving online learning experience. Previous studies about online learning technologies have focused on students' acceptance of the technology \cite{allen2003videoconferencing}, the effectiveness and equity of the knowledge delivery \cite{koceski2013challenges, lawson2010images}, or pedagogical benefits and constraints \cite{davis2019online, gillies2008student}. During the pandemic, much research has been conducted to understand online education in this new context, including examining the benefits and challenges of remote teaching and learning \cite{bergdahl2020covid, dhawan2020online}, or summarizing views and attitudes of teachers and students \cite{adnan2020online, akyildiz2020college, hebebci2020investigation}. Some recent studies have touched upon VC techniques and affordances, but they either refer generally to overall online interaction \cite{hacker2020virtually, parkbeyond}, or focus on contexts other than learning, for example, online working and team collaboration \cite{waizenegger2020affordance}. Questions regarding how learners use technology remain unanswered, such as: why do students opt for different affordances of the VC platforms, and how does the use of specific affordances affect their learning experience? The present study bridges this gap by proposing a general research question that asks how the usage of videoconferencing affordances affects students’ remote learning experience.


Based on the theory of interactive media effects (TIME) \cite{sundar2015toward}, the present study attempts to answer the research question by looking at three types of affordances of VC technologies: modality (specifically, the use of video and audio), interactivity (the use of chat and raising-hand function), and agency (customization). As we are interested in whether the use of these affordances would make a difference to students’ learning experience online, we posed a series of specific questions, as follows:  

1) Modality question: Is it better for students and their classmates to use the video function? By showing their faces, the communication may be more organic as it approximates the way they communicate in person where they can observe visual cues from other people. However, it is not clear if this will lead to a more engaged learning experience, or end up feeling too invasive to be comfortable for learning. After all, even in the physical classroom, not all students are looking at one student. Compared to teachers, who may feel more at home delivering their instruction to a wall full of live student images rather than black screens with names, students may prefer less invasive ways of showing their presence, like participating via chat or the raise-hand function. On the other hand, if they see that other students are using video, will that contribute to their own class experience and encourage them to turn on their own video camera? 

2) Interactivity question: How does interaction play a role in virtual classroom? Interaction has always been seen as an indispensable component of knowledge acquisition and cognitive development in face-to-face (FtF) education \cite{song2011understanding}. Offline instructors enjoy a highly communicative environment. They make an effort to recognize each student, build immediacy by showing verbal and nonverbal behaviors, and to create interactivity in the classroom that makes students feel comfortable. VC techniques try to afford the same things by providing various communication channels. But do students really appreciate being recognized? Or, does the appeal of VC depend on the absence of such features, given past findings about students enjoying online courses due to their asynchronicity, anonymity, and flexibility \cite{song2004improving}? 

3) Agency question: Is greater user control helpful or distracting in the virtual learning context? As students were forced to move to online classes, they may have felt less in control of their learning. Can the VC platform help them to gain more control in class interaction by allowing them to customize their account or interface? Although students cannot choose the mode of learning, they can choose how they present themselves to others while they are learning, which could be quite empowering. Alternatively, they might feel the customization options are unnecessary and distract them from engaging in the process of learning.

Answers to such questions and concerns will be valuable in understanding how technology can shape the future of remote education and provide pointers on best practices for designing and using them effectively.

\section{Theories and Hypotheses}
Considering the common features incorporated in VC software, we examine psychological mechanisms of affordance use in this section by reviewing prior theories and research in the field of computer-mediated communication (CMC), which addresses how individuals behave and subtly shape communications in a virtual relational context \cite{walther2011theories}, as we develop a set of hypotheses and research questions for study.

\subsection{Affordance}

The concept of “affordance” was coined by Gibson to describe the actionable properties between an environment and its actors \cite{gibson1977theory}. Norman applied it to product design, and emphasized that affordances should be perceivable \cite{don1988design}. Maier and Fadel proposed an alternative approach that refers to affordances as a special set of interactions between designers, artifacts, and users \cite{maier2009affordance}. Within such designer-artifact-user complex systems, affordances describe potential behaviors between two or more subsystems \cite{maier2009affordance}. Based on this logic, Maier and Fadel came up with the framework for affordance-based design. Applying this to the domain of media, Sundar et al. suggested that affordances are manifested in the form of unique technological attributes that shape the nature of communication \cite{sundar2015toward}. According to their theory of interactive media effects (TIME), affordances can affect user psychology by triggering actions or by serving as cues on the interface \cite{sundar2015toward}.  In this paper, we focus on the interactive nature of affordances and consider videoconferencing affordances as possibilities offered by certain VC features regarding how to interact and communicate with others. We focus on three types of affordances of VC technologies used in TIME, specifically modality, interactivity, and agency affordance, which are embedded in the common features found on most VC platforms.

\subsection{The self-usage of affordances}

Videoconferencing techniques support knowledge delivery and in-class communication in remote learning by providing visual, audio, and other interactive affordances, and attempt to transcend social distance while maintaining the synchronicity of classroom interactions. These affordances are achieved by features like video camera, audio, chat function, and so forth. Furthermore, many VC interfaces allow users to customize their account, such as changing the name and photo shown in the class and using a virtual background during the meeting. These customized options help students to shape their personalities and identities \cite{sundar2008self}, thus enabling them to control class interaction. To benefit maximally from VC affordances, the usage of a variety of features is essential \cite{saykili2018distance}. In fact, the use of technology is a cornerstone in the field of distance learning \cite{alharthi2020students}. The affordances provided by VC platforms shape the landscape of how participants communicate and collaborate in class. With rich possibilities for computer-mediated class communication about course content, students' learning experiences can be vastly enhanced. We therefore propose:

\begin{displayquote}
\textbf{H1}: Student’s self-usage of VC affordances (video, audio, chat, raise-hand, and customization features) will be positively related to their class (a) engagement, (b) interaction, and (c) satisfaction.
\end{displayquote}

\subsection{Social conformity}
Social learning theory suggests that learning is a social process, and thus highly influenced by our surroundings \cite{bandura1977social}. This social environment also applies to the virtual world \cite{smith2009social}, as students keep an eye on their classmates in order to adjust self-behaviors during class meetings. Social conformity theory suggests a change of behavior due to group influence, resulting in congruence between individuals and group \cite{beran2015conformity, wijenayake2020impact}. In an online class, if many of their classmates have video cameras on, students are likely to conform and turn on their own video regardless of their original intentions. Once such social environment is established and sensed by students, they might feel more engaged, interactive, and satisfied. Therefore:

\begin{displayquote}
\textbf{H2}: Students’ perception of others’ usage of the video camera feature in class will be positively related to their own class (a) engagement, (b) interaction, and (c) satisfaction.
\end{displayquote}

\begin{displayquote}
\textbf{H3}: Students’ self-usage of the video camera feature will be positively related to their perception of others’ usage of video camera in class.
\end{displayquote}

\subsection{Mediators}

\subsubsection{Perceived anonymity}
Anonymity is a critical factor that affects students’ willingness to participate in class exercises \cite{freeman2006anonymity}. To avoid social risks in learning, students are able to obtain security from anonymity \cite{barr2017encouraging}. The Social Identity Model of Deindividuation Effects (SIDE) theory argues that users who communicate via computers without seeing each other experience visual anonymity, leading to greater interpersonal anonymity \cite{walther2015interpersonal}. With VC software, visual anonymity can be easily achieved by turning off the video camera. Furthermore, discursive anonymity \cite{scott2004benefits} can be achieved by avoiding verbal communicative channels such as audio and chat. Together, these two types of anonymity contribute to the overall sense of interpersonal anonymity in communication. 

Students favor anonymity, in part because it can help to reduce the impact of social norms and allow them to interact fully and equally, devoid of the social constraints in FtF communication \cite{flanagin2002computer}. Many studies also associate anonymity with higher self-disclosure and responsiveness, more collaboration and participation in learning \cite{barr2017encouraging, chester1998online, clark2019anonymity, freeman2004student}. On the other hand, students also like FtF interaction when they want to build interpersonal relationships \cite{paechter2010online}; therefore, reducing anonymity is a good way to approximate FtF experiences. During the pandemic, since students are going online out of necessity, they may indeed be eager to develop interpersonal relationships with one another. Hence, we posit that reducing anonymity would be positively related to students’ experience:

\begin{displayquote}
\textbf{H4}: The use of video, audio, and chat features will be negatively associated with students' perceived anonymity and positively related to their class (a) engagement, (b) interaction, and (c) satisfaction.
\end{displayquote}

\subsubsection{Nonverbal cues}
The Social Information Processing (SIP) theory \cite{walther2015interpersonal} suggests that even though conventional non-verbal cues are common in face-to-face communications, they may not present in online settings; individuals compensate for them by perceiving and conveying communicative messages through a combination of cues. In the remote classroom context, students sense emotions of instructors and classmates through many nonverbal cues such as facial expressions, gestures, and head nodding from video streaming \cite{han2013nonverbal, walther2015interpersonal}. These nonverbal behaviors show presence, express support, forecast response, and convey nuanced sentiments \cite{parkbeyond}, and thus lead to more affective and immediate interactions \cite{mehrabian1971silent}. Therefore, we propose:

\begin{displayquote}
\textbf{H5}: Other students’ usage of video cameras will be positively associated with perceived nonverbal cues and therefore positively related to class (a) engagement, (b) interaction, and (c) satisfaction.
\end{displayquote}

\subsubsection{Comfort level}
Past evidence indicates that students perform better in learning when they feel comfortable \cite{wilson2001contributing}. They are also more willing to engage in the class, given reduced fear and resistance. “Comfort” was labeled as a successful mediating factor in blended learning by some researchers. For instance, in Futch et al.'s interview with instructors, a FtF classroom was deemed a more comfortable place to start the course \cite{futch2016comfort}. On the other hand, there are studies that suggest remote courses mediated by technology are better because online classrooms help create a comfortable environment for informal learning of certain topics, like foreign languages \cite{ngo2019college, soltovets2019foreign}. When many students have their cameras on, a multi-screen projection in a VC-based class may give students an approximate sense of FtF communication \cite{szeto2016towards}, and thus make them comfortable during class interactions. Additionally, as an easy-to-observe behavioral engagement in class, students raise their hands before responding to a topic or before interrupting other speakers \cite{boheim2020student}. Such behavior, which is less intrusive, could give students a higher level of comfort. As Futch et al. note, students sometimes feel embarrassed about raising hands in the FtF classroom but feel safer in the online environment \cite{futch2016comfort}.

Based on these arguments, we propose the following hypotheses:

\begin{displayquote}
\textbf{H6}: Other students’ usage of video cameras will be positively associated with comfort level and therefore positively related to class (a) engagement, (b) interaction, and (c) satisfaction.
\end{displayquote}

\begin{displayquote}
\textbf{H7}: Students' usage of the raise-hand function will be positively associated with their comfort level and therefore positively related to their class (a) engagement, (b) interaction, and (c) satisfaction.
\end{displayquote}

The diagram below (Figure 1) summarizes the hypotheses. The independent variables (IVs) on the left-hand side of the diagram refer to the three sets of affordances in the form of five features, with video and audio referring to modality affordance, chat and raising hand referring to interactivity affordance, and customization referring to agency affordance. The effects of self-usage of modality features on the dependent variables (DVs) of class engagement, interaction and satisfaction are hypothesized by H1 (total effect) and H4 (mediation via anonymity) whereas the effects of their usage by classmates are hypothesized by H2 (total effect) and H5 (mediation via nonverbal cues) and H6 (mediation via comfort level). The relationship between others’ usage and self usage of video is explored by H3. The proposed effects of self-usage of interactivity features (chat and raise hand) on the DVs are articulated by H1 (total effect), H4 (mediation via anonymity) and H7 (mediation via comfort level). The proposed effect of the agency feature of customization on the DVs is embedded in H1. Together, this study model investigates the psychological effects of self-usage of the three sets of affordances and other-usage of the modality affordance of video camera upon learners’ online class experience.

\begin{figure}[h!]
  \caption{Summary of Hypotheses}
  \includegraphics[width=\textwidth]{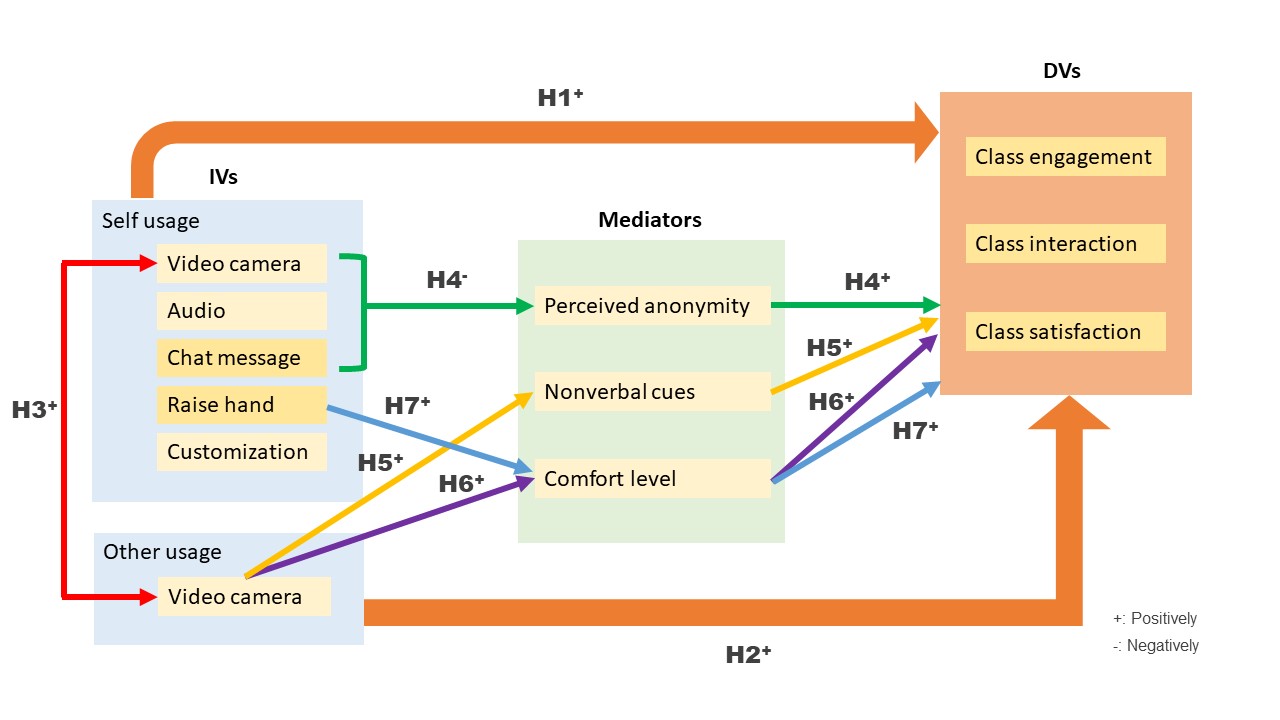}
\end{figure}




\subsection{Control variables}
\subsubsection{Interest and Autonomy}
When students show a high level of interest in a certain class, they tend to have positive evaluations \cite{anwar2019developing,naceur2005motivation, schiefele1999interest}. Meanwhile, autonomous students engage in learning affectively, because they feel a sense of ownership and responsibility for their class work \cite{lee2015autonomy}. Autonomy is also involved in the learner-centered teaching method, where students actively take part in collaboration and participation \cite{yoon2020predictive}. Both concepts are associated with intrinsic motivation for learning \cite{naceur2005motivation, serdyukova2013student}, meaning that self-driven students might be more inclined to use VC affordances in class, or they are more likely to be engaged and satisfied. This study controlled for interest and autonomy in order to ensure that affordances usage was not simply an artifact of differences in the students' initial engagement with their classes.

\subsubsection{Gender and Class size}
Substantial literature exists on the effects of gender on online behaviors. For example, early studies revealed that women are more inclined to conform because they are more easily swayed by external factors than men \cite{eagly1986sex}. More recent empirical studies show that individuals are sensitive to gender cues in online settings, which leads to higher conformity through triggering stereotypical perceptions of the competency of peers \cite{wijenayake2019measuring}. Gender also has effects on participation in online classrooms, where females felt more connected to their peers, and the online learning experiences were more aligned with their educational goals \cite{rovai2005gender}. Social presence and interactivity in classes were moderated by gender difference; to be specific, male students benefited more from interactive technologies in enhancing presence and increasing satisfaction with online classes \cite{park2020exploring}. 

Class size is also known to affect learning experience \cite{blatchford2020rethinking}. Studies examining the effects of class size on computer-supported collaborative leaning found that teaching quality in large classes was limited \cite{yang2020stage}. Peer relations, interactions and influences were stronger in smaller classes \cite{olagbaju2019effects}. Therefore, we include these two factors (gender and class size) in our study as statistical controls.

\section{Method}

The study was approved by the Institutional Review Board (IRB). An online survey was conducted via Qualtrics as Study 1, between April 23, 2020 and May 6, 2020, within the first two months of the pandemic, aiming to capture reactions of students to the novel situation in a timely fashion. Participants were recruited through three channels: 1) Email distribution and snowball sampling through school connections; 2) Research Participation System provided by the university, wherein participants were granted extra course credits for taking the survey; and 3) Amazon’s Mechanical Turk platform, where users can request or perform human intelligence tasks (HIT) in return for payment. To ensure quality responses, participants needed to have a HIT approval rate of greater than or equal to 96\% to be qualified for this study. They were paid 60 cents each for participation, given that the mean estimated time for completing the survey was less than 10 minutes.

\subsection{Participants}

All participants in Study 1 were U.S. students and were taking at least one remote synchronous class at the start of the lockdown period. Since we are interested in students' experience in general, we did not ask the respondents to report their educational class but only to specify whether they are undergraduate or graduate students.

200 individuals filled out the survey. After excluding those answering in less than 30 seconds and those who failed the attention check questions, the final sample had 176 responses. Among the 176 participants, 50.6\% were female (N = 89)  and 49.4\% were male (N = 87). The mean age was 24.74 years old (SD = 6.53). 60.8\% of them were undergraduates and 36.4\% were in graduate programs. 85.2\% of the participants reported Zoom as their major tool for taking the remote classes. 40.91\% of the respondents came from email distribution, 22.16\% were from the university’s Research Participation System, and 36.93\% were from MTurk.

\subsection{Measures}

Participants in Study 1 were asked to think about one of their most recent remote class meetings and fill out the questionnaire. The usage of VC affordances was divided into several dimensions: students' self-usage of the video camera feature during the class, other students’ usage of video cameras in the same class, students' self-usage of audio, then chat and raise hand functions compared to other students, and students' self-usage of customization features. Considering the variance of features that different VC tools would provide, for each question the participants had the option of choosing “not applicable”.

Video and audio usage were measured by asking participants what percentage of the time (from 0\% to 100\%) they turned their camera (M = 52.13\%, SD = 39.31) or audio (M = 47.07\%, SD = 38.22) on during the whole class, and what percentage of their classmates turned on cameras (M = 47.79\%, SD = 36.03). We divided the responses into a higher usage group and a lower usage group based on a median split. For the usage of chat (M = 3.83, SD = 1.40) and raise hand (M = 3.16, SD = 1.50) functions, we asked participants to compare themselves with classmates and indicate whether they used these functions more frequently or less frequently than others on 7-point scales, ranging from “much less” to “much more” with a middle point being “about the same”. Data were coded into three groups: lower usage (score < 4), average usage (score = 4), and higher usage (score > 4).

Then, participants were asked to specify the ways in which they customized their interface for the remote class, including changing their profile photo, renaming themselves, using virtual background in their video, and other operations provided by certain VC tools.  We provided options such as “others” and “none” to avoid forcing choices. Given the nature of the variance in the resulting data, this variable was re-coded dichotomously because we are interested in whether customization influences the experience rather than the differential effects of different types of customization. Participants who had used one or more customization options were placed in the customization group, with the remaining in the non-customization group.

The items measuring dependent variables were adapted from existing scales and modified to fit the remote learning context. Students’ class engagement was measured by the extent to which they took active steps to maintain their class performance, and the extent to which they intend to be involved in the class \cite{dixson2015measuring, gunuc2015student}. Class interaction includes the interaction between the participants and the instructors, and the interaction between the participants and their classmates \cite{sherry1998assessing}. Class satisfaction reflects how much they learned from the class, how much the class meets their expectations and how likely they would be to recommend this class to others \cite{gunawardena1997social, sher2009assessing}. Perceived anonymity was conceptualized as how visible and identifiable they felt to the instructor and their classmates. Nonverbal cues refer to the facial expressions and body language of classmates that participants observed in class \cite{mottet2000interactive}. Comfort level reflects how comfortable students felt when participating in class activities \cite{gunawardena1997social, kim2011developing}. Interest indicates to what extent participants found their class content interesting, and autonomy suggests how they control their learning \cite{walker2005development}. All these measures were assessed on 7-point Likert-type scales ranging from 1 (strongly disagree) to 7 (strongly agree), except for the interest scale, which ranged from “not interesting at all” to “extremely interesting.” Table 1 displays survey items, their pedigree, psychometric properties and descriptives in our sample.

\begin{table}
  \scriptsize
  \centering
  \caption{Measurement Items in Study 1}
  \label{tab:freq}
  \begin{tabular}{p{2.2cm} p{1.1cm} p{0.6cm} p{0.6cm} p{35em} }
    \toprule
   Measured variable & Cronbach’s $\alpha$ & Mean & SD & Items\\
    \midrule 
    Class engagement & .85 & 5.16 & 1.02 &  “I prepared in advance before the class meeting”; “I have been regular with my homework for this class”; “I have been taking notes during the class”; “I feel anxious when I don’t attend the class”; “I am an active student in class”; “I attend the class willingly”; “I try to find ways to make the course material relevant to my life”; “I carefully listen to the instructor and other students during the class”; “I try to do my best regarding my responsibilities in group work”; and “I try to help other students in this class”.\\
    Class interaction & .86 & 4.50 & 1.30 &  “The level of interaction between all participants is high”; “In general, the instructor is effective in motivating the students to interact in class”; “Compared to other students, I often state my opinions to the instructor”; “Compared to other students, I often answer the instructor’s questions”; “Compared to other students, I often ask questions in the class”; and “I interacted with other students a lot”. \\
    Class satisfaction & .88 & 5.17 & 1.16 &  “I was able to learn from the content of this remote class”; “I was able to learn from the class discussion”; “I was stimulating to do additional reading or research on topics discussed in class”; “Overall, the class meets my expectation”; “Overall, I am satisfied with this class”; and “I would recommend this course to other students”. \\
    Perceived anonymity & .60 & 2.89 & 1.28 &  “When I was talking, I felt the instructor and other students were looking at me”; “I believe the instructor and other students could recognize me when I was talking”; and “I believe the instructor and other students could recognize me when I send a message on chat”.\\
    Nonverbal cues & .81 & 4.47 & 1.46 &  “I often saw other students’ gestures when they were talking in the class”;  “I often saw the other students nodding their heads while I was talking to them”; and “I often saw the other students varied their facial expressions to show feelings when they were talking”. \\
    Comfort levels & .80 & 5.06 & 1.22 &  “I felt comfortable conversing in class through this medium”; “I felt comfortable participating in class discussions”; and “I felt comfortable interacting with other participants in the class”. \\
    Interest & N/A & 5.34 & 1.25 &  “To what extent do you find the content of the class (i.e., the topics, not instructor or mode of instruction) interesting?” \\
    Autonomy & .79 & 5.54 & 1.00 &  “As a student, I make decisions about my learning”; “I am in control of my learning”; and “I do course-related work during times that I find convenient”.  \\
  \bottomrule
\end{tabular}
\end{table}

Aside from Cronbach's $\alpha$, we checked composite reliability and discriminant validity by running a confirmatory factor analysis (CFA). The results showed a reasonable fit of the model to the data: $\chi^2$ = 948.95,  \textit{df} = 419, \textit{p} <.001, 95\%CI [.08, .09], RMSEA = .08 , CFI = .80 \cite{browne1992alternative}. All factor loadings were significant (\textit{p}’s < .01). The results also confirmed that composite reliabilities (CR) for each construct exceeded the minimum criteria of .70. Discriminant validity was assessed with Fornell and Larcker’s criterion \cite{fornell1981structural} of checking whether the average variance extracted (AVE) for each construct exceeds the square of the pairwise correlations between constructs. All AVE estimates were greater than the squared inter-construct correlations, thereby confirming discriminant validity, i.e., distinctness of the measures. This means they are not collinear.

\section{Results}

The correlations among the key variables are reported in Table 2.

\begin{table}
  \tiny
  \caption{Pearson Correlations Among Key Variables in Study 1}
  \label{tab:freq}
  \centering
  \begin{tabular}{m{5.5em} m{0.7cm} m{0.7cm} m{0.7cm} m{0.7cm}  m{0.7cm} m{0.7cm} m{0.7cm} m{0.7cm} m{0.7cm} m{0.7cm} m{0.7cm} m{0.7cm}}
    \toprule
    Variables & 1 & 2 & 3 & 4 & 5 & 6 & 7 & 8 & 9 & 10 & 11 & 12 \\
    \midrule 
    1 Self video usage & 1 &&&&&&&&&&&\\
    2 Other video usage & .588** & 1 &&&&&&&&&&\\
    3 Self audio usage & .330** & .301** & 1 &&&&&&&&&\\
    4 Self chat usage & -.023 & .026 & -.084 & 1 &&&&&&&&\\
    5 Self raise-hand usage & .149 & .172* & .061 & .193*  & 1 &&&&&&\\
    6 Customization & -.005 & .058 & .253** & -.027  & .124 & 1 &&&&&&\\
    7 Perceived anonymity & -.422** & -.398** & -.199* & -.227**  & -.139 & -.129 & 1 &&&&&\\
    8 Nonverbal cues & .370** & .293** & .209* & -.031  & .103 & .127 & -.428** & 1 &&&&\\
    9 Comfort level & .286** & .313** & .171* & .234**  & .198* & .085 & -.413** & .449** & 1 &&&\\
    10 Class engagement & .336** & .267** & .195* & .186*  & .115 & -.080 & -.355** & .379** & .513** & 1 &&\\
    11 Class interaction & .338** & .310** & .238** & .234**  & .292** & .149 & -.405** & .533** & .619** & .577** & 1 &\\
    12 Class satisfaction & .256** & .275** & .191* & .212**  & .245** & .055 & -.421** & .469** & .651** & .582** & .667** & 1\\
  \bottomrule
  \multicolumn{7}{l}{*p$<$.05, **p$<$.01}  \\
\end{tabular}
\end{table}

\subsection{Main effects}
A series of analyses of covariance (ANCOVAs) were conducted to test the main effects, with interest, autonomy, gender, and class size as covariates (see Table 3). No significant difference was found between undergraduate and graduate students.

\begin{table}
\renewcommand\arraystretch{1.3}
  \scriptsize
  \caption{ANCOVA Test for Main Effects in Study 1 }
  \label{tab:freq}
    \centering
  \begin{tabular}{m{3cm} m{1cm} m{0.5cm} m{1cm} m{3cm} m{1cm} m{2cm}}
    \toprule
     &  & M & SE & F & {\textit{p}} & partial $\eta^2$\\
    \midrule 
     & \multicolumn{6}{c}{Class engagement}\\
        \cline{2-7}
     \multirow{2}*{Self-video usage} & High & 5.39 & .12 & \multirow{2}*{\textit{F} (1, 129) = 5.61} & \multirow{2}*{.019} & \multirow{2}*{.04}\\
		~ & Low & 4.95 & .13 & ~ & ~ & ~\\ 
		\cline{2-7}
	 \multirow{2}*{Self-audio usage} & High & 5.37 & .11 & \multirow{2}*{\textit{F} (1, 134) = 8.08} & \multirow{2}*{.005} & \multirow{2}*{.06}\\
	~ & Low & 4.92 & .12 & ~ & ~ & ~\\ 
	    \cline{2-7}
	\multirow{3}*{Self-chat usage} & High & 5.29 & .15 & \multirow{3}*{\textit{F} (2, 133) = 2.03} & \multirow{3}*{.136} & \multirow{3}*{.03}\\
	~ & Average & 5.27& .13 & ~ & ~ & ~\\
	~ & Low & 4.94 & .14 & ~ & ~ & ~\\ 
		\cline{2-7}
	\multirow{3}*{Self raise-hand usage} & High & 5.26 & .21 & \multirow{3}*{\textit{F} (2, 133) = .21} & \multirow{3}*{.809} & \multirow{3}*{.00}\\
	~ & Average & 5.19& .13 & ~ & ~ & ~\\
	~ & Low & 5.11 & .12 & ~ & ~ & ~\\ 
	   \cline{2-7}
     \multirow{2}*{Customization} & Yes & 5.15 & .10 & \multirow{2}*{\textit{F} (1, 133) = .03} & \multirow{2}*{.857} & \multirow{2}*{.00}\\
		~ & No & 5.18 & .13 & ~ & ~ & ~\\ 
	   \cline{2-7}
     \multirow{2}*{Other-video usage} & High & 5.35 & .11 & \multirow{2}*{\textit{F} (1, 131) = 5.27} & \multirow{2}*{.023} & \multirow{2}*{.04}\\
		~ & Low & 4.98 & .12 & ~ & ~ & ~\\
	\midrule 
	& \multicolumn{6}{c}{Class interaction}\\
        \cline{2-7}
     \multirow{2}*{Self-video usage} & High & 4.75 & .15 & \multirow{2}*{\textit{F} (1, 145) = 2.91} & \multirow{2}*{.090} & \multirow{2}*{.02}\\
		~ & Low & 4.35 & .16 & ~ & ~ & ~\\ 
		\cline{2-7}
	 \multirow{2}*{Self-audio usage} & High & 4.83 & .13 & \multirow{2}*{\textit{F} (1, 150) = 10.33} & \multirow{2}*{.002} & \multirow{2}*{.06}\\
	~ & Low & 4.21 & .14 & ~ & ~ & ~\\ 
	    \cline{2-7}
	\multirow{3}*{Self-chat usage} & High & 4.82 & .18 & \multirow{3}*{\textit{F} (2, 150) = 2.16} & \multirow{3}*{.120} & \multirow{3}*{.03}\\
	~ & Average & 4.48& .16 & ~ & ~ & ~\\
	~ & Low & 4.31 & .17 & ~ & ~ & ~\\ 
		\cline{2-7}
	\multirow{3}*{Self raise-hand usage} & High & 5.30 & .24 & \multirow{3}*{\textit{F} (2, 150) = 6.15} & \multirow{3}*{.003} & \multirow{3}*{.08}\\
	~ & Average & 4.41& .15 & ~ & ~ & ~\\
	~ & Low & 4.35 & .14 & ~ & ~ & ~\\ 
	   \cline{2-7}
     \multirow{2}*{Customization} & Yes & 4.68 & .13 & \multirow{2}*{\textit{F} (1, 150) = 4.06} & \multirow{2}*{.046} & \multirow{2}*{.03}\\
		~ & No & 4.28 & .15 & ~ & ~ & ~\\ 
	   \cline{2-7}
     \multirow{2}*{Other-video usage} & High & 4.81 & .13 & \multirow{2}*{\textit{F} (1, 147) = 7.93} & \multirow{2}*{.006} & \multirow{2}*{.05}\\
		~ & Low & 4.26 & .14 & ~ & ~ & ~\\
	\midrule 
	& \multicolumn{6}{c}{Class satisfaction}\\
        \cline{2-7}
     \multirow{2}*{Self-video usage} & High & 5.32 & .12 & \multirow{2}*{\textit{F} (1, 148) = .77} & \multirow{2}*{.383} & \multirow{2}*{.01}\\
		~ & Low & 5.15 & .13 & ~ & ~ & ~\\ 
		\cline{2-7}
	 \multirow{2}*{Self-audio usage} & High & 5.42 & .11 & \multirow{2}*{\textit{F} (1, 153) = 7.53} & \multirow{2}*{.007} & \multirow{2}*{.05}\\
	~ & Low & 4.98 & .11 & ~ & ~ & ~\\ 
	    \cline{2-7}
	\multirow{3}*{Self-chat usage} & High & 5.39 & .15 & \multirow{3}*{\textit{F} (2, 153) = 2.63} & \multirow{3}*{.076} & \multirow{3}*{.03}\\
	~ & Average & 5.29& .13 & ~ & ~ & ~\\
	~ & Low & 4.95 & .13 & ~ & ~ & ~\\ 
		\cline{2-7}
	\multirow{3}*{Self raise-hand usage} & High & 5.65 & .21 & \multirow{3}*{\textit{F} (2, 153) = 3.62} & \multirow{3}*{.029} & \multirow{3}*{.05}\\
	~ & Average & 5.23& .13 & ~ & ~ & ~\\
	~ & Low & 5.01 & .12 & ~ & ~ & ~\\ 
	   \cline{2-7}
     \multirow{2}*{Customization} & Yes & 5.28 & .10 & \multirow{2}*{\textit{F} (1, 153) = 1.89} & \multirow{2}*{.171} & \multirow{2}*{.01}\\
		~ & No & 5.06 & .13 & ~ & ~ & ~\\ 
	   \cline{2-7}
     \multirow{2}*{Other-video usage} & High & 5.42 & .12 & \multirow{2}*{\textit{F} (1, 150) = 6.09} & \multirow{2}*{.015} & \multirow{2}*{.04}\\
		~ & Low & 5.02 & .12 & ~ & ~ & ~\\
  \bottomrule
\end{tabular}
\end{table}

Results for Study 1 showed that students with high usage of video cameras in class felt more engaged than those who had low usage. Students who have high usage of audio also felt more engaged, and interacted more and expressed higher satisfaction. No significant relationship emerged for chat usage. Compared to those who had lower and average usage, students who had higher usage of the raise-hand function than their classmates had a better experience in terms of interaction and satisfaction. Students who customized their accounts interacted more in class compared to those who did not. Therefore, H1 was partially supported. When students perceived their classmates used the video feature more often in the class, they reported a higher level of engagement, interaction, and satisfaction compared to their counterparts who perceived lesser number of classmates using video. H2 was therefore supported.

A 2 x 2 chi-square test was conducted to examine H3 pertaining to the relationship between other-usage and self-usage of video camera. Data revealed a significantly larger proportion of students who felt "many" classmates had cameras on would also use video themselves than those said few of their classmates turned on the camera, $\chi^2$ (1, N = 165) = 56.95, \textit{V}* = .59, \textit{p} < .001. Specifically, when classmates had a higher usage of video in class, 80\% of the respondents reported that they also had high usage of cameras, while only 21.3\% students reported that they had high video usage when they felt "few" classmates doing so. Hence, H3 was supported.

\subsection{Mediation effects}

To examine the hypothesized mediating roles of perceived anonymity (H4), perceived nonverbal cues (H5), and comfort level (H6, H7), a series of simple mediation analysis (Model 4 of PROCESS macro) \cite{hayes2017introduction} was conducted, using 5,000 bootstrap samples and 95\% bias-corrected confidence intervals (see Table 4).

\begin{table}
\renewcommand\arraystretch{1.3}
  \scriptsize
  \caption{Indirect Effects of Affordances Usage on DVs via Mediators in Study 1}
  \label{tab:freq}
    \centering
  \begin{tabular}{m{2cm} m{2.5cm} m{1.5cm} m{1.5cm} m{1cm} m{1cm}}
    \toprule
    \multicolumn{3}{l}{\multirow{2}*{Mediation Path}} & \multicolumn{1}{c}{\multirow{2}*{Indirect effect Bootstrap Estimate(b)}} & \multicolumn{2}{c}{Indirect effect 95\% C.I.}\\
    &&&& LL & UL\\
    \midrule 
    \multirow{3}*{Self-video usage $\rightarrow$} & \multirow{3}*{Perceived Anonymity $\rightarrow$} & Engagement & .10 (.07) & -.01 & .26\\
    ~ & ~ & Interaction & .11 (.07) & -.01 & .27\\
    ~ & ~ & Satisfaction & .11 (.07) & .00 & .28\\
    \midrule 
    \multirow{3}*{Self-audio usage $\rightarrow$} & \multirow{3}*{Perceived Anonymity $\rightarrow$} & Engagement & .12 (.07) & .01 & .28\\
    ~ & ~ & Interaction & .14 (.08) & .02 & .34\\
    ~ & ~ & Satisfaction & .13 (.07) & .02 & .29\\
    \midrule 
    \multirow{3}*{Self-chat usage $\rightarrow$} & \multirow{3}*{Perceived Anonymity $\rightarrow$} & Engagement & .06 (.05) & -.01 & .18\\
    ~ & ~ & Interaction & .08 (.05) & -.00 & .19\\
    ~ & ~ & Satisfaction & .06 (.04) & -.01 & .16\\
    \midrule 
    \multirow{3}*{Other-video usage $\rightarrow$} & \multirow{3}*{Nonverbal Cues $\rightarrow$} & Engagement & .17 (.07) & .05 & .31\\
    ~ & ~ & Interaction & .27 (.11) & .08 & .52\\
    ~ & ~ & Satisfaction & .18 (.08) & .05 & .36\\
    \midrule 
    \multirow{3}*{Other-video usage $\rightarrow$} & \multirow{3}*{Comfort Level $\rightarrow$} & Engagement & .21 (.09) & .07 & .40\\
    ~ & ~ & Interaction & .34 (.10) & .15 & .55\\
    ~ & ~ & Satisfaction & .25 (.09) & .08 & .44\\
    \midrule 
    \multirow{3}*{Raise hand usage $\rightarrow$} & \multirow{3}*{Comfort Level $\rightarrow$} & Engagement & .10 (.06) & .00 & .22\\
    ~ & ~ & Interaction & .14 (.08) & -.00 & .31\\
    ~ & ~ & Satisfaction & .12 (.06) & .00 & .26\\
       
  \bottomrule
\end{tabular}
\end{table}

For H4, results showed that higher usage of the audio function is related to reduced perceived anonymity, which in turn was positively associated with class-related outcomes. Higher self-video usage was associated with lower perceived anonymity, which in turn was positively related to class satisfaction. No significant indirect effects via perceived anonymity were found for class engagement or interaction, though the data patterns were similar. The indirect effects of self-chat usage via perceived anonymity on class engagement, interaction, or satisfaction failed to achieve statistical significance, although it was in the same direction. Therefore, H4 was partially supported.

For H5, results showed that when classmates had a higher level of video usage, students perceived more nonverbal cues, which was related to greater class engagement, interaction, and satisfaction. Thus, H5 was supported. Similarly, comfort produced by other-video usage was associated with better experiences, H6 was supported. Additionally, raised-hand usage was associated with a higher comfort level and thus related to greater class satisfaction. H7 was partially supported.

\section{Discussion}

To understand how the usage of VC affordances influences students’ remote learning experience, we conducted survey during the early stage of the transition to online education, which captured the timely reactions from students regarding their behaviors and perceptions towards learning remotely via VC techniques. While there are many actors in educational activities, we centered on students' experience given their main role in the learning practice. These experiences largely rely on how they use the affordances of VC tech, as it becomes the main force in supporting remote education now and probably in the future.

\subsection{Summary of results }
Table 5 summarizes the supported hypothesis testing results. 

\begin{table}
\renewcommand\arraystretch{1.3}
  \scriptsize
  \caption{Summary of the Supported Hypothesis}
  \label{tab:freq}
    \centering
  \begin{tabular}{m{1cm} m{10cm}}
    \toprule
    & Supported hypotheses\\
    \midrule 
    H1a & more self-usage of video $\rightarrow$ greater class engagement\\
     & more self-usage of audio $\rightarrow$ greater class engagement\\
    H1b & more self-usage of audio  $\rightarrow$ greater class interaction\\
     & more self-usage of hand-raising  $\rightarrow$ greater class interaction\\
     & customization $\rightarrow$ greater class interaction\\
    H1c & more self-usage of audio  $\rightarrow$ greater class satisfaction\\
    & more self-usage of hand-raising  $\rightarrow$ greater class satisfaction\\
    H2a & more other-usage of video $\rightarrow$ greater class engagement\\
    H2b & more other-usage of video $\rightarrow$ greater class interaction\\
    H2c & more other-usage of video $\rightarrow$ greater class satisfaction\\
    H3 & more other-usage of video $\rightarrow$ more self-usage of video\\
    H4a & more self-usage of audio $\rightarrow$ less perceived anonymity $\rightarrow$ greater class engagement\\
    H4b & more self-usage of audio $\rightarrow$ less perceived anonymity $\rightarrow$ greater class interaction\\
    H4c & more self-usage of video $\rightarrow$ less perceived anonymity $\rightarrow$ greater class satisfaction\\
     & more self-usage of audio $\rightarrow$ less perceived anonymity $\rightarrow$ greater class satisfaction\\
    H5a & more other-usage of video $\rightarrow$ more nonverbal cues $\rightarrow$ greater class engagement\\
    H5b & more other-usage of video $\rightarrow$ more nonverbal cues $\rightarrow$ greater class interaction\\
    H5c & more other-usage of video $\rightarrow$ more nonverbal cues $\rightarrow$ greater class satisfaction\\
    H6a & more other-usage of video $\rightarrow$ higher comfort level $\rightarrow$ greater class engagement\\
    H6b & more other-usage of video $\rightarrow$ higher comfort level $\rightarrow$ greater class interaction\\
    H6c & more other-usage of video $\rightarrow$ higher comfort level $\rightarrow$ greater class satisfaction\\
    H7a & more self-usage of hand-raising $\rightarrow$ higher comfort level $\rightarrow$ greater class engagement\\
    H7c & more self-usage of hand-raising $\rightarrow$ higher comfort level $\rightarrow$ greater class satisfaction\\
  \bottomrule
\end{tabular}
\end{table}
Our results indicated a positive correlation between the usage of VC affordances and students’ class experience, which answers the general research question in this study. Specifically, students felt less anonymous with their use of video and audio functions, which was related to a positive class experience. When their classmates turned on their cameras, students felt the class as more engaging, interactive, and satisfying. The raise hand function was positively associated with students' feeling of comfort and contributed to their class engagement as well as satisfaction. Students had higher class interaction when they customized their accounts. We also discovered two psychological mechanisms: 1) Anonymity does not appear to be helpful in the remote class context because we found that the usage of video and audio functions contributes to positive class experience by reducing the sense of anonymity; 2) Classmates’ video usage positively influences students’ class reactions by conveying nonverbal cues, increasing comfort level, and encouraging self-usage of video. 

These findings have strong theoretical and practical implications for an affordance-based approach to technology-supported communication and learning that is based on psychological mechanisms derived from CMC literature. While our hypotheses received overwhelming support from the data, it is not clear if the effects are due to the novelty of the situation, since all data were collected soon after the transition to online classes from offline classes, in the middle of the semester in Spring, 2020. Therefore, before going further into interpreting the findings and drawing larger implications, we conducted a constructive replication of this study well after the novelty of the transition had worn off, after making some refinements that are described in the next section.

\subsection{Study 2}
The second survey was distributed on February 20, 2021, and concluded on May 7, 2021, which covered much of the spring semester of 2021, allowing students sufficient time to respond. We had a larger sample size for this study compared to Study 1. To minimize sampling errors, we narrowed down the sampling frame. Participants were only recruited from a public university in the eastern United States. While email distribution and the snowball sampling method through school connections were also used, participants were recruited mainly through the same Research Participation System that we used in the first survey. In addition, the phrasing of survey items was refined to further reduce any possible confusion for the respondents.

\subsubsection{Participants}
Same as Study 1, all participants in Study 2 were U.S. students and were taking at least one remote synchronous class. 303 responses were received, and 256 responses were used in the final data analysis, after excluding those who did not complete the main part of the questionnaire and those who indicated that they were taking only asynchronous courses.  Among the 256 responses, 69.14\% were female (N = 177) and 30.86\% were male (N = 79). The mean age was 21.42 years old (SD = 3.56). 90.63\% of them were undergraduates and 8.98\% were in graduate program. 
98.05\% of the participants reported Zoom as their major tool for taking remote classes. 22.66\% of the respondents were from email distribution, and 77.34\% came from the Research Participation System.

\subsubsection{Measures}
Study 2 inherited the variables from Study 1 with some wording refinement on certain items (Table 6). The measures for students’ self-usage of video cameras (M = 24.29\%, SD = 34.19) and audio (M = 16.77\%, SD = 27.98) during the class remained the same, but we specified their observation of classmates’ usage of video by asking what percentage of students have their video cameras on (M = 20.87\%,SD = 29.37) during the class and what percentage of the time during the class period when other students have their video cameras on (M = 34.53\%, SD = 36.92). These two dimensions will give participants an overall sense of the extent to which their classmates tend to use videos. Data was coded into the combination of these two measures by multiplying them (M = .14, SD = .26). These responses were used to create a high usage group and low usage group based on a median split. In addition, we specified the chat usage by differentiating public chat from private chat, since they only can observe others who use public chat in class (M = 3.42, SD = 1.74).  The measure for virtual hand usage remained the same (M = 2.87, SD = 1.78). Data were coded into three groups:  lower usage (score < 4), average usage (score = 4), and higher usage (score > 4).  The measure and code for customization remained the same.

\begin{table}
  \scriptsize
  \caption{Measurement Items in Study 2}
  \label{tab:freq}
    \centering
  \begin{tabular}{p{2.2cm} p{1.1cm} p{0.6cm} p{0.6cm} p{35em} }
    \toprule
   Measured variable & Cronbach’s $\alpha$ & Mean & SD & Items)\\
    \midrule 
    Class engagement & .82 & 4.99 & 1.00 &  “I prepared in advance before the class meeting”; “I have been regular with my homework for this class”; “I have been taking notes during the class”; “I feel anxious when I don’t attend the class”; “I am an active student in class”; “I attend the class willingly”; “I try to find ways to make the course material relevant to my life”; “I carefully listen to the instructor and other students during the class”; and “I try to do my best regarding my responsibilities in group work”.\\
    Class interaction & .86 & 3.37 & 1.34 &  “The level of interaction between all participants is high”; “Compared to other students, I often state my opinions to the instructor”; “Compared to other students, I often answer the instructor’s questions”; “Compared to other students, I often ask questions in the class”; and “I interacted with other students a lot”. \\
    Class satisfaction & .92 & 4.82 & 1.36 &  “I was able to learn from the content of this remote class”; “I was able to learn from the class discussion”; “Overall, I am satisfied with this class”; and “I would recommend this course to other students”. \\
    Perceived anonymity & .88 & 3.76 & 1.63 & “I believe the instructor and other students could recognize me when I was talking”; “I believe the instructor and other students could recognize me when I send a message in chat”; and "I felt identifiable in the class".\\
    Nonverbal cues & .81 & 4.27 & 1.05 &  "I often saw other students raised their hands to talk"; "I often saw other students using vocal starters that suggested they wanted to say something"; "I often saw the instructor gesture with hands or arms when he/she was talking in the class"; “I often saw other students gesture with hands or arms when they were talking in the class”; "I often saw the instructor nodding his/her head while I was talking to him/her"; “I often saw the other students nodding their heads while I was talking to them”; "I often saw the instructor varied his/her facial expressions to show attitude or feelings when he/she was talking"; and “I often saw the other students varied their facial expressions to show attitude or feelings”. \\
    Comfort levels & .85 & 4.77 & 1.28 &  “I felt comfortable communicating with the instructor or other students by speaking”; "I felt comfortable communicating with the instructor or other students by text, using the chat function"; “I felt comfortable participating in class discussions, either by audio or chat”; “I felt comfortable interacting with other participants in the class”; and "I felt comfortable learning through this online format". \\
    Interest & N/A & 4.68 & 1.43 &  “To what extent do you find the content of the class (i.e., the topics, not instructor or mode of instruction) interesting?” \\
    Autonomy & .79 & 5.64 & .89 &  “As a student, I make decisions about my learning”; “I am in control of my learning”; “I do course-related work during times that I find convenient”; "I play an important role in my learning"; and "I approach learning in my own way".  \\
  \bottomrule
\end{tabular}
\end{table}

\subsubsection{Main effects}
\begin{table}
  \tiny
  \caption{Pearson Correlations Among Key Variables in Study 2}
  \label{tab:freq}
    \centering
  \begin{tabular}{m{5.5em} m{0.7cm} m{0.7cm} m{0.7cm} m{0.7cm}  m{0.7cm} m{0.7cm} m{0.7cm} m{0.7cm} m{0.7cm} m{0.7cm} m{0.7cm} m{0.7cm}}
    \toprule
    Variables & 1 & 2 & 3 & 4 & 5 & 6 & 7 & 8 & 9 & 10 & 11 & 12\\
    \midrule 
    1 Self video usage & 1 &&&&&&&&&&&\\
    2 Other video usage & .695** & 1 &&&&&&&&&&\\
    3 Self audio usage & .243** & .132* & 1 &&&&&&&&&\\
    4 Self chat usage & .087 & -.037 & .170** & 1 &&&&&&&&\\
    5 Self raise-hand usage & .275** & .140* & .217** & .406** & 1 &&&&&&&\\
    6 Customization & .112 & .022 & .034 & .129* & .152*  & 1 &&&&&&\\
    7 Perceived anonymity & -.369** & -.225** & -.120 & -.263** & -.304** & -.175** & 1 &&&&&\\
    8 Nonverbal cues & .315** & .361** & .195** & .161* & .215** & .140* & -.348 & 1 &&&&\\
    9 Comfort level & .295** & .243** & .248** & .352** & .274** & .124* & -.562** & .429**  & 1 &&&\\
    10 Class engagement & .289** & .176** & .077 & .192** & .108 & .056 & -.280** & .291** & .434**  & 1 &&\\
    11 Class interaction & .425** & .225** & .261** & .376** & .310 & .218** & -.548** & .434** & .643** & .452** & 1 &\\
    12 Class satisfaction & .232** & .195** & .042 & .201** & .134* & .024 & -.315** & .389** & .567**  & .571** & .500** & 1\\
  \bottomrule
  \multicolumn{7}{l}{*p$<$.05, **p$<$.01}  \\
\end{tabular}
\end{table}

\begin{table}
\renewcommand\arraystretch{1.3}
  \scriptsize
  \caption{ANCOVA Test for Main Effects in Study 2 }
  \label{tab:freq}
    \centering
  \begin{tabular}{m{3cm} m{1cm} m{0.5cm} m{1cm} m{3cm} m{1cm} m{2cm}}
    \toprule
     &  & M & SE & F & {\textit{p}} & partial $\eta^2$\\
    \midrule 
     & \multicolumn{6}{c}{Class engagement}\\
        \cline{2-7}
     \multirow{2}*{Self-video usage} & High & 5.10 & .08 & \multirow{2}*{\textit{F} (1, 247) = 2.91} & \multirow{2}*{.090} & \multirow{2}*{.01}\\
		~ & Low & 4.89 & .08 & ~ & ~ & ~\\ 
		\cline{2-7}
	 \multirow{2}*{Self-audio usage} & High & 5.12 & .08 & \multirow{2}*{\textit{F} (1, 248) = 5.02} & \multirow{2}*{.026} & \multirow{2}*{.02}\\
	~ & Low & 4.87 & .08 & ~ & ~ & ~\\ 
	    \cline{2-7}
	\multirow{3}*{Self-chat usage} & High & 5.28 & .11 & \multirow{3}*{\textit{F} (2, 247) = 6.13} & \multirow{3}*{.003} & \multirow{3}*{.05}\\
	~ & Average & 5.04& .09 & ~ & ~ & ~\\
	~ & Low & 4.79 & .08 & ~ & ~ & ~\\ 
		\cline{2-7}
	\multirow{3}*{Self raise-hand usage} & High & 5.28 & .15 & \multirow{3}*{\textit{F} (2, 247) = 2.36} & \multirow{3}*{.097} & \multirow{3}*{.02}\\
	~ & Average & 4.98& .10 & ~ & ~ & ~\\
	~ & Low & 4.92 & .07 & ~ & ~ & ~\\ 
	   \cline{2-7}
     \multirow{2}*{Customization} & Yes & 4.98 & .08 & \multirow{2}*{\textit{F} (1, 248) = .05} & \multirow{2}*{.820} & \multirow{2}*{.00}\\
		~ & No & 5.00 & .08 & ~ & ~ & ~\\ 
	   \cline{2-7}
     \multirow{2}*{Other-video usage} & High & 5.23 & .11 & \multirow{2}*{\textit{F} (1, 248) = 5.87} & \multirow{2}*{.016} & \multirow{2}*{.02}\\
		~ & Low & 4.91 & .06 & ~ & ~ & ~\\
	\midrule 
	& \multicolumn{6}{c}{Class interaction}\\
        \cline{2-7}
     \multirow{2}*{Self-video usage} & High & 3.80 & .12 & \multirow{2}*{\textit{F} (1, 247) = 22.91} & \multirow{2}*{.000} & \multirow{2}*{.09}\\
		~ & Low & 2.98 & .11 & ~ & ~ & ~\\ 
		\cline{2-7}
	 \multirow{2}*{Self-audio usage} & High & 4.04 & .10 & \multirow{2}*{\textit{F} (1, 248) = 77.73} & \multirow{2}*{.000} & \multirow{2}*{.24}\\
	~ & Low & 4.21 & .14 & ~ & ~ & ~\\ 
	    \cline{2-7}
	\multirow{3}*{Self-chat usage} & High & 3.91 & .16 & \multirow{3}*{\textit{F} (2, 247) = 12.02} & \multirow{3}*{.000} & \multirow{3}*{.09}\\
	~ & Average & 3.54& .13 & ~ & ~ & ~\\
	~ & Low & 2.97 & .12 & ~ & ~ & ~\\ 
		\cline{2-7}
	\multirow{3}*{Self raise-hand usage} & High & 3.85 & .21 & \multirow{3}*{\textit{F} (2, 247) = 7.17} & \multirow{3}*{.001} & \multirow{3}*{.06}\\
	~ & Average & 3.64& .14 & ~ & ~ & ~\\
	~ & Low & 3.11 & .11 & ~ & ~ & ~\\ 
	   \cline{2-7}
     \multirow{2}*{Customization} & Yes & 3.59 & .11 & \multirow{2}*{\textit{F} (1, 248) = 6.31} & \multirow{2}*{.013} & \multirow{2}*{.03}\\
		~ & No & 3.18 & .11 & ~ & ~ & ~\\ 
	   \cline{2-7}
     \multirow{2}*{Other-video usage} & High & 3.83 & .16 & \multirow{2}*{\textit{F} (1, 248) = 9.85} & \multirow{2}*{.002} & \multirow{2}*{.04}\\
		~ & Low & 3.24 & .09 & ~ & ~ & ~\\
	\midrule 
	& \multicolumn{6}{c}{Class satisfaction}\\
        \cline{2-7}
     \multirow{2}*{Self-video usage} & High & 4.94 & .11 & \multirow{2}*{\textit{F} (1, 247) = 1.82} & \multirow{2}*{.178} & \multirow{2}*{.01}\\
		~ & Low & 4.72 & .10 & ~ & ~ & ~\\ 
		\cline{2-7}
	 \multirow{2}*{Self-audio usage} & High & 4.97 & .10 & \multirow{2}*{\textit{F} (1, 248) = 3.56} & \multirow{2}*{.057} & \multirow{2}*{.01}\\
	~ & Low & 4.70 & .10 & ~ & ~ & ~\\ 
	    \cline{2-7}
	\multirow{3}*{Self-chat usage} & High & 4.96 & .14 & \multirow{3}*{\textit{F} (2, 247) = 3.22} & \multirow{3}*{.042} & \multirow{3}*{.03}\\
	~ & Average & 5.00 &.12 & ~ & ~ & ~\\
	~ & Low & 4.62 & .11 & ~ & ~ & ~\\ 
		\cline{2-7}
	\multirow{3}*{Self raise-hand usage} & High & 5.11 & .18 & \multirow{3}*{\textit{F} (2, 247) = 1.39} & \multirow{3}*{.251} & \multirow{3}*{.01}\\
	~ & Average & 4.81& .13 & ~ & ~ & ~\\
	~ & Low & 4.76 & .09 & ~ & ~ & ~\\ 
	   \cline{2-7}
     \multirow{2}*{Customization} & Yes & 4.75 & .10 & \multirow{2}*{\textit{F} (1, 248) = 1.09} & \multirow{2}*{.298} & \multirow{2}*{.00}\\
		~ & No & 4.90 & .10 & ~ & ~ & ~\\ 
	   \cline{2-7}
     \multirow{2}*{Other-video usage} & High & 5.00 & .14 & \multirow{2}*{\textit{F} (1, 248) = 1.84} & \multirow{2}*{.176} & \multirow{2}*{.01}\\
		~ & Low & 4.77 & .08 & ~ & ~ & ~\\
  \bottomrule
\end{tabular}
\end{table}

As can be seen from the Table 8, the results of Study 2 followed the same pattern as Study 1. Although some tests in Study 2 failed to achieve statistical significance, data were all in the same direction. Tests with chat usage were more significant compared to Study 1. Students who used the chat function more in class reported a higher level of engagement, interaction, and satisfaction. Students who had a higher usage of all kinds of affordance and who perceived their classmates used more video felt more interaction in the classroom.

Data manifested the relationship between self-use of camera and classmates' use of cameras in Study 2 was consistent. Students (85.5\%) who perceived their classmates used the video feature more in class tended to use video camera more compared to students (37.1\%) who felt less video usage from their classmates $\chi^2$ (1, N = 256) = 44.00, \textit{V}* = .42, \textit{p} < .001.

\subsubsection{Mediation effects}
Results for mediation effects in Study 2 were consistent with those in Study 1 (see Table 9). 

In sum, Study 2 results were consistent with those found in Study 1, with some achieving more significant results.

\begin{table}
\renewcommand\arraystretch{1.3}
  \scriptsize
  \caption{Indirect Effects of Affordances Usage on DVs via Mediators in Study 2}
  \label{tab:freq}
    \centering
  \begin{tabular}{m{2cm} m{2.5cm} m{1.5cm} m{1.5cm} m{1cm} m{1cm}}
    \toprule
    \multicolumn{3}{l}{\multirow{2}*{Mediation Path}} & \multicolumn{1}{c}{\multirow{2}*{Indirect effect Bootstrap Estimate(b)}} & \multicolumn{2}{c}{Indirect effect 95\% C.I.}\\
    &&&& LL & UL\\
    \midrule 
    \multirow{3}*{Self-video usage $\rightarrow$} & \multirow{3}*{Perceived Anonymity $\rightarrow$} & Engagement & .11 (.04) & .04 & .19\\
    ~ & ~ & Interaction & .30 (.08) & .15 & .47\\
    ~ & ~ & Satisfaction & .14 (.05) & .05 & .23\\
    \midrule 
    \multirow{3}*{Self-audio usage $\rightarrow$} & \multirow{3}*{Perceived Anonymity $\rightarrow$} & Engagement & .12 (.04) & .05 & .22\\
    ~ & ~ & Interaction & .29 (.07) & .16 & .44\\
    ~ & ~ & Satisfaction & .15 (.06) & .06 & .27\\
    \midrule 
    \multirow{3}*{Self-chat usage $\rightarrow$} & \multirow{3}*{Perceived Anonymity $\rightarrow$} & Engagement & .05 (.02) & .02 & .09\\
    ~ & ~ & Interaction & .15 (.05) & .06 & .24\\
    ~ & ~ & Satisfaction & .07 (.03) & .02 & .12\\
    \midrule 
    \multirow{3}*{Other-video usage $\rightarrow$} & \multirow{3}*{Nonverbal Cues $\rightarrow$} & Engagement & .12 (.04) & .04 & .22\\
    ~ & ~ & Interaction & .31 (.08) & .16 & .48\\
    ~ & ~ & Satisfaction & .23 (.07) & .11 & .40\\
    \midrule 
    \multirow{3}*{Other-video usage $\rightarrow$} & \multirow{3}*{Comfort Level $\rightarrow$} & Engagement & .13 (.04) & .05 & .22\\
    ~ & ~ & Interaction & .34 (.10) & .14 & .54\\
    ~ & ~ & Satisfaction & .26 (.08) & .11 & .42\\
    \midrule 
    \multirow{3}*{Raise hand usage $\rightarrow$} & \multirow{3}*{Comfort Level $\rightarrow$} & Engagement & .10 (.03) & .04 & .16\\
    ~ & ~ & Interaction & .24 (.06) & .13 & .35\\
    ~ & ~ & Satisfaction & .19 (.05) & .10 & .29\\
       
  \bottomrule
\end{tabular}
\end{table}

\subsection{Implications for Technology-Supported Communication and Learning}

From a theoretical standpoint, our data questions current understanding of the role of anonymity in virtual interactions and motivates us to take more seriously the social aspects of affordance usage in such settings.

\subsubsection{Reconsider anonymity} Our findings extend our understanding of the role played by anonymity in the context of online learning. While previous studies found that anonymity can give confidence and encourage students to take part in online discussion \cite{freeman2004student} by overcoming the fear of criticism \cite{caspi2008media}, our study suggests that, in synchronous remote classes, the affordances that make students less anonymous (such as using video and audio) are in fact associated with greater engagement, interaction, and class satisfaction. This could be interpreted as a concerted effort to maintain the interpersonal relationships which may have been developed before. Unlike traditional online courses, where students rarely interact with each other, remote class during the pandemic inherits the pre-existing connections among students. They appreciate being recognized and appreciate approximating the FtF communication pattern. By turning off the video and audio, perceived social presence may be reduced for students’ interaction with classmates \cite{aragon2003creating}. Social presence is important for maintaining immediacy \cite{gunawardena1997social} and is also linked to more satisfaction and perceived learning \cite{andel2020social}. Being less socially present may be perceived as being less available and involved \cite{park2015can}. It is also possible that anonymity can lead to a feeling of reduced accountability and encourage students to become “lurkers” that fail to contribute content or participate \cite{wilkerson2016lurking}. Regardless of turning on the video and audio voluntarily or based on request, students obtain a positive experience from it by getting more involved and feeling more present. To better understand and utilize anonymity in promoting student’s learning, the goals, context, and preexisting status should be taken into consideration. Anonymity is not always desirable for achieving higher student satisfaction.

\subsubsection{Importance of social aspects} Students had a positive class experience when more of their classmates turned on their video cameras. As SIP theory addresses, an individual can decode nonverbal cues from various online channels. In the synchronous remote context, nonverbal behaviors observed via the video feeds appeared to have made the conversations more organic and contributed to students’ comfort level with class activities.

In addition, the study showed that participants were more inclined to have their cameras on when their classmates had them on. One possible reason is that they may reciprocate with their appearance as equal contribution to the interaction. Such mimicry behavior has been documented by many researchers (for reviews, see \cite{chartrand2002you, chartrand1999chameleon}). Individuals tend to behave similarly with their communicative partners, and in many cases, they do so unconsciously. Another reason could be social conformity, which describes an act of changing one’s behavior to match that of others. Like FtF class meetings, students in the remote class may observe their classmates’ behavioral norms, and adjust their own behaviors accordingly to maintain a positive self-concept \cite{cialdini2004social}. Moreover, social norms also seem to apply to some self-initiated actions such as raising a virtual hand before expressing ideas when other people are talking. While the current pedagogical culture encourages free, equal and spontaneous participation in the classroom, the norm of “raise your hand to speak” is ingrained in the American culture \cite{dixon2009they}. In FtF communication, people occasionally step in the conversation and accommodate each other naturally. However, in the online synchronous context, the risk of speaking over one another is higher due to differential transmission of audio signals. By raising their virtual hands in the VC interface, students may feel that they follow good manners and show respect to other speakers. All these mutually reinforcing behaviors appear to contribute to students’ sense of comfort, enhancing their engagement with learning and satisfaction with the class. These findings emphasize the social attributes in a remote context. With the trend of a “new normal” or a mixed mode in education, educators and learners should pay more attention to these social aspects and make use of them to enhance students’ learning experience. They also shed light on the importance of comfort level in students’ learning. Either by increasing nonverbal cues in the class, or by offering consistency in transferring FtF social norms to the online context, VC affordances serve to make students comfortable in remote learning.

\subsection{Implications for Designing Supportive Digital Tools}

To take advantage of VC tech in remote learning and collaboration, teachers and designers should encourage the use of affordances, especially those related to visibility or identity. For example, video streaming is an important affordance -- students not only benefit from turning on their own cameras, but also benefit from others’ adoption. It provides nonverbal cues and helps approximate FtF communication, augmenting their sense of immediacy and social presence. Therefore, teachers may encourage students to use video feed s during class, and designers should devote themselves to solving technical issues such as bandwidth, compatibility, and resolution. They could also develop new video functions to intrigue users, such as creating a virtual classroom scene that allows students to incorporate their video images, giving them a sense of immersion. This could pave the way for designing a metaverse environment for virtual learning.

Additionally, educators should be aware of the social aspects in remote classes, paying particular attention to promoting their comfort level. Unlike asynchronous online courses, remote learning at this time and possibly mixed-mode learning in the future both require interactions and social connections among class participants. This also can be seen as a collaboration on learning activities. Besides raising virtual hands, designers could add more functions to support the norms of interpersonal and group communications. For example, they could explore ways to allow students to assess others’ reactions when they are presenting in class, so that they feel like they are speaking to a human audience rather than a machine, perhaps by affording eye contact between the presenter and listeners. Furthermore, customization is also a good way to get students involved as our findings showed greater class interactions when students customized their accounts. In addition to the current customization options of self-presentation (e.g. customizing a virtual background or account profile), VC designers could develop features that allow users to adjust their interface layouts to see other people’s features, and change how they want their class interfaces to look (e.g., hide others’ video window, arrange the order of the video window, or be able to separate the chat box and move it to a desired location).

\subsection{Limitations and future study}

A major limitation of the two studies is the relatively small sample size. Furthermore, our recruiting methods and the use of convenience samples may have lowered the reliability of the responses. However, we did find several significant results from the small sample, and the effects could become more salient with a larger sample. Additionally, since year of education is not our focus, we did not include this information in the surveys. Our results found no significant differences between undergraduate students and graduate students. Future research can further differentiate students' year of study to see if any patterns can be found. Furthermore, both studies relied on a survey method that, while externally valid, is limited in its ability to establish causal relationships. Future research in a more controlled experimental setting is required to formally test the theoretical mechanisms suggested by our significant indirect effects. Another limitation is that, our surveys only explored the most common affordances for examination, as most of our participants indicated Zoom as their main tool in remote learning. Future studies should examine other affordances as many new VC platforms have rapidly emerged to meet the sudden need for scalable online classes; for example, shared creation affordance provided by whiteboard, or gamification affordance provided by UI animation. In addition, there could be plenty of legitimate reasons other than conformity that explain why students use video functions or other functions in class. Future work could explore their usage more deeply by adopting qualitative methods such as in-depth interviews.

\section{Conclusion}

The role of technology in facilitating learning and communication has attracted scholars’ attention for decades, with distance education in particular highlighting the importance of online learning technology. The availability of software and affordances is vital, as they can become powerful tools for remote learning to approximate the experience of in-person education. Our studies took advantage of the social distancing period, measured students’ experience during an unexpected transition, and obtained many valuable insights for the design of educational technology. Notwithstanding some limitations, our studies make a significant contribution by shining the spotlight on common affordances of VC platforms (modality, interactivity, and agency) and identifying key mediators (perceived anonymity, nonverbal cues, and comfort level) by which their use (both by respondents themselves and their peers) is positively related to class engagement, interaction and satisfaction with online learning experience.

\bibliographystyle{style.bst}  
\bibliography{references}

\end{document}